\documentclass[letterpaper]{JHEP3}

\usepackage{amsmath,amsthm, amssymb}
\usepackage{epsfig,multicol}

\newcommand{\be}{\begin{equation}}
\newcommand{\ee}{\end{equation}}
\newcommand{\ba}{\begin{array}}
\newcommand{\ea}{\end{array}}
\newcommand{\bea}{\begin{eqnarray}}
\newcommand{\eea}{\end{eqnarray}}
\newcommand{\bma}{\begin{matrix}}
\newcommand{\ema}{\end{matrix}}
\newcommand{\bpm}{\begin{pmatrix}}
\newcommand{\epm}{\end{pmatrix}}
\newcommand{\nn}{\nonumber}
\newcommand{\half}{\textstyle{\frac{1}{2}}}
\newcommand{\dst}{\displaystyle}
\newcommand{\tst}{\textstyle}
\newcommand{\mc}{\mathcal}
\newcommand{\mb}{\mathbb}

\newcommand{\p}{\partial}
\newcommand{\hf}{{\hat5}}

\newcommand{\ov}{\overline}
\newcommand{\wh}{\widehat}
\newcommand{\wt}{\widetilde}
\newcommand{\Psitil}{\wt \Psi}

\newcommand{\etabar}{\ov \eta}

\newcommand{\cP}{{\bf P}}
\newcommand{\bP}{{{\bf P}_{{}_{tw}}}}

\newcommand{\eps}{\epsilon}
\newcommand{\al}{\alpha}

\newcommand{\la}{\lambda}
\newcommand{\da}{\delta}
\newcommand{\om}{\omega}
\newcommand{\Ga}{\Gamma}
\newcommand{\Si}{\Sigma}
\newcommand{\si}{\sigma}
\newcommand{\vp}{\varphi}
\newcommand{\vt}{\vartheta}
\newcommand{\Mink}{{\it Mink$_4$}}
\newcommand{\AdS}{{\it AdS$_4$}}

\title{Twisting Warped Supergravity}
\author{Jonathan Bagger {\it and} Dmitry Belyaev\\ 
Department of Physics and 
Astronomy, The Johns Hopkins University, \\
3400 North Charles Street, Baltimore, MD 21218, USA\\ 
E-mail: \email{bagger@jhu.edu}\\
E-mail: \email{belyaev@pha.jhu.edu}}

\preprint{}

\abstract{
We study gauged five-dimensional supergravity on the interval
$[0,\pi R]$.  We find a set of boundary conditions with respect to which
the theory is locally supersymmetric.  For theories with detuned
brane tensions ($\Lambda_4<0$), we show that these boundary
conditions can be used to spontaneously break global supersymmetry.
For the original, tuned Randall-Sundrum scenario ($\Lambda_4=0$),
we prove that the locally supersymmetric boundary conditions
are also globally supersymmetric.  We lift the theory from $[0,\pi R]$
to $S^1$ and $\mb{R}$, with arbitrary twists for the fermions, and
cast these results in the language of the Scherk-Schwarz
mechanism.
}


\begin{document}


\section{Introduction}

The supersymmetric Randall-Sundrum scenario \cite{rs,old,bb1} is based on five-dimensional 
supergravity on a manifold with boundary.  The fifth dimension is the interval
$[0,\pi R]$, usually realized as the orbifold $S^1/\mathbb{Z}_2$.  
The background is warped, with a nontrivial profile in the fifth dimension.  
The bulk contains pure supergravity with a cosmological constant 
$\Lambda_5=-6\la^2$; the boundaries correspond to three-branes with 
tensions $T_0=6\la_0$ and $T_\pi=-6\la_\pi$.

In a recent paper \cite{bb1}, we constructed the locally supersymmetric action
for the ``detuned" case, in which $T_0$, $T_\pi$ and $\Lambda_5$ are not
related.  We found that supersymmetry imposes boundary conditions on the
bulk fields and requires $|\la_{0,\pi}|\leq\la$.  The background was warped, with
either a flat (\Mink) or anti-de Sitter (\AdS) metric on the four-dimensional
slices.  The theory was locally supersymmetric whenever the boundary conditions
were satisfied.  In this paper, we identify a subset of the boundary conditions
for which supersymmetry is spontaneously broken.

We begin by considering warped supergravity in the ``downstairs'' picture,
in which the fifth dimension is the interval  $[0,\pi R]$.  This approach is
self-consistent and has the advantage that it avoids certain complications 
(discontinuous fields, delta-function singularities, etc.) that arise in the 
``upstairs'' picture, on $S^1$ or $\mb{R}$.  In the downstairs picture,
the Scherk-Schwarz mechanism \cite{ss} is nothing but spontaneous
supersymmetry breaking by boundary conditions.  We will show that
this breaking can be accomplished in every \AdS\ background, but never
in the case of \Mink.

We then lift  $[0,\pi R]$ to the ``upstairs" picture, in which the fifth
dimension is $S^1$ or $\mb{R}$.  We use a {\it broken} symmetry
to construct a ``twisted" lifting, in which the fermions rotate along
the extra dimension, while the bosons remain periodic.  The method
is very general and can be used in other applications.  

In the ``upstairs'' picture, the fields can twist and jump.  
There is freedom to interchange the twists and the jumps,
maintaining the same boundary conditions on the fundamental domain.  
We emphasize that all liftings describe the same physics.

\section{Working ``downstairs''}

\subsection{Introduction}

In this section we summarize the results from Ref.~\cite{bb1},
converting them to the ``downstairs'' picture, in which the fifth
dimension is the interval $z\in[0,\pi R],$  rather than the circle
$S^1$.  We then derive a new result:  in the {\it AdS$_4$} warped background,
supersymmetry can be spontaneously broken by a {\it continuous}
family of boundary conditions.  In the next section, we will see
how the supersymmetry breaking can also be described by a
Scherk-Schwarz twist.

\subsection{Bulk action}
\label{babc}

The bulk action of gauged five-dimensional supergravity is
\bea
\label{bulk}
S_{bulk} \ =\  \int d^5\!x e_5 
\Big\{
-\frac{1}{2} R 
+\frac{i}{2}\Psitil_M^i \Ga^{MNK} D_N \Psi_{Ki} 
-\frac{1}{4}F_{MN}F^{MN} 
\nn\\
+ 6\la^2 
-\frac{3}{2} \la\, \vec q\cdot \vec\si_i{}^j \Psitil_M^i \Si^{MN}  \Psi_{Nj} 
+\dots
\Big\}.
\eea
The action depends on a unit vector $\vec q=(q_1, q_2, q_3)$ that
defines a gauged $U_{\vec q}(1)$ subgroup of the $SU(2)_R$
automorphism group.
(The complete action, its supersymmetry transformations, and our
conventions are collected in Ref.~\cite{bb1}.)

The action of an element $U\in SU(2)_R$ on a symplectic Majorana
spinor $\Psi_i$ (and on its two-component constituents $\psi_i$)
is given by
\be
\Psi_i^\prime=\wt U_i{}^j\Psi_j, \quad
\psi_i^\prime=U_i{}^j\psi_j,
\ee
where $\wt U =\si_3 U\si_3$.
The $SU(2)_R$ symmetry is broken by the gauging.  The symmetry
is formally restored if we rotate the parameters as follows,
\be
(\vec q{\,}^\prime\cdot\vec\si) = 
\wt U (\vec q\cdot\vec\si) \wt U^\dagger, \quad
(\vec p{\,}^\prime\cdot\vec\si) = U (\vec p\cdot\vec\si) U^\dagger.
\ee
Here $\vec p=(-q_1, -q_2, q_3)$ is defined by $(\vec p\cdot\vec\si)
=\si_3 (\vec q\cdot\vec\si)\si_3$.  The unbroken $U_{\vec q}(1)$
subgroup is given by
\be
\wt U=\exp(i\omega\vec q\cdot\vec \si), \quad
U=\exp(i\omega\vec p\cdot\vec \si).
\ee
%

\subsection{Boundary conditions}

Under (local) supersymmetry, variation of the action (\ref{bulk}) 
gives rise a boundary term:
\be
\label{bvar}
\da S_{bulk} = \int_{\mathcal{M}} d^5\!x (\p_M K^M) =
\int_{\p\mathcal{M}} d^4\!x (n_M K^M) = 
\int_{z=\pi R} d^4\!x K^5 - \int_{z=0} d^4\!x K^5.
\ee
In this section we present boundary conditions that make
the boundary term vanish.  

We start by imposing the simplest bosonic boundary conditions 
that support warped backgrounds.  On the boundary at $z=0$,
we take
\be
\label{bbc0}
e_m^\hf \ =\  e_5^a\ =\ B_m\ =\ 0, \quad
\om_{ma\hf} \ =\ \la_0 e_{ma}.
\ee
Demanding that these conditions be preserved under
supersymmetry gives rise to the following fermionic
boundary conditions:
\bea
\label{fbc0}
\eta_2 \ =\  \al_0\eta_1, \quad
\psi_{m2}\ =\ \al_0\psi_{m1}, \quad 
\psi_{51} \ =\ -\al_0^\ast\psi_{52}.
\eea
Supersymmetry also requires that the parameters $\la_0$ and $\al_0$
be related as follows,
\be
\label{fff}
\la_0=f(\al_0, \vec q\;)\la, \quad
f(\al, \vec q\;) = 
-\frac{(\al+\al^\ast)q_1+i(\al^\ast-\al)q_2+(\al\al^\ast-1)q_3}
{1+\al\al^\ast}.
\ee
This relation couples the fermionic and bosonic boundary
conditions.  The boundary conditions decouple in the flat case, 
when $\la=0$. 

For the boundary at $z=\pi R$, identical reasoning gives an analogous
set of boundary conditions:
\bea
\label{bbc1}
&&e_m^\hf \ =\  e_5^a\ =\ B_m\ =\ 0, \quad
\om_{ma\hf} \ =\ \la_\pi e_{ma}, \\
\label{fbc1}
&&\eta_2 \ =\  \al_\pi\eta_1, \quad
\psi_{m2}\ =\ \al_\pi\psi_{m1}, \quad 
\psi_{51} \ =\ -\al_\pi^\ast\psi_{52},
\eea
where
\be
\label{fbc2}
\la_\pi = f(\al_\pi, \vec q\;) \la.
\ee
The parameters $\al_0$ and $\al_\pi$ are not related because
the supersymmetry is local.  The full set of boundary
conditions forms a two-parameter family.

With the above boundary conditions, the 
action is invariant under $N=2$
supersymmetry, with arbitrary supersymmetry parameters
$\eta_1(x,z)$ and $\eta_2(x,z)$ in the bulk, restricted only
on the boundary:
\be
\label{etabc}
\eta_2(x, 0) = \al_0 \eta_1(x, 0), \quad
\eta_2(x, \pi R) = \al_\pi \eta_1(x, \pi R).
\ee
Note that in the ``downstairs'' picture, there is no boundary action.
Brane actions appear only on the covering space, where
the parameters $T_0=6\la_0$ and $T_\pi=-6\la_\pi$ play
the role of brane tensions.

\subsection{Warped backgrounds}

A bosonic background, consistent with the equations of motion 
and the boundary conditions, is given by
\be
e_m^a=a(z)\wh e_m^a(x), \quad e_5^\hf=1, \quad
e_m^\hf=e_5^a=B_m=B_5=0.
\ee
The warp factor $a(z)$ satisfies the equation
\be
a^\prime(z)^2\ =\ \la^2(a(z)^2-K^2).
\ee
The boundary conditions follow from $\om_{ma\hf}=
-(a^\prime/a)\,e_{ma}$,
\be
a^\prime(0)=-\la_0\,a(0), \quad
a^\prime(\pi R)=-\la_\pi\,a(\pi R).
\ee
With this ansatz, the four-dimensional veirbein $\wh e_m^a(x)$
solves Einstein's equations with cosmological constant 
$\Lambda_4=-3\la^2 K^2$.  The normalization
condition $a(0)=1$ fixes $K$ and $a(z)$ uniquely.

Local supersymmetry requires  $|\la_{0,\pi}|\leq\la$
(as follows from $|f(\al, \vec q\;)|\leq 1$), so there are three
distinct cases:
\begin{enumerate}
\item[(1)]
$-\la<\la_\pi<\la_0<\la$.  This gives 
$a(z)=K\cosh(\la z-c_0)$, where 
\be
K=\sqrt{1-\left(\frac{\la_0}{\la}\right)^2}, \quad
c_0=\frac{1}{2}\log\left(\frac{\la+\la_0}{\la-\la_0}\right).
\ee
The metric on the four-dimensional slices is \AdS, with $\Lambda_4<0$;
the distance $R$ is fixed:
\be
\label{rrr}
\la\pi R =\frac{1}{2}\log
\left(\frac{\la+\la_0}{\la-\la_0}\cdot
\frac{\la-\la_\pi}{\la+\la_\pi}\right).
\ee
\item[(2)]
$\la_\pi=\la_0=\pm\la$.  This is the Randall-Sundrum case, with
$a(z)=\exp(\mp\la z)$, and flat four-dimensional slices, 
\Mink, with $\Lambda_4=0$.  The distance $R$ is arbitrary.
\item[(3)]
$\la_\pi\neq \la_0=\pm\la$.  This is the case where one brane
tension is tuned to the bulk cosmological constant, but the other
is not.  In this case there is no static background.
\end{enumerate}

\subsection{Supersymmetry breaking}

A bosonic background is (globally) supersymmetric if there is
a solution to the Killing spinor equations:
\be
\da\psi_{m1,2}=0, \quad \da\psi_{51,2}=0.
\ee
For the case at hand, they are
\bea
2\wh D_m\eta_1+i\om_{ma\hf}\si^a\etabar_2+i\la\si_m
(q_3\etabar_2+q_{12}^\ast\etabar_1)=0, \quad
2\p_5\eta_1+\la(q_3\eta_1-q_{12}^\ast\eta_2)=0,
\nn\\
2\wh D_m\eta_2-i\om_{ma\hf}\si^a\etabar_1+i\la\si_m
(q_3\etabar_1-q_{12}\etabar_2)=0, \quad
2\p_5\eta_2-\la(q_3\eta_2+q_{12}\eta_1)=0,
\eea
where $q_{12}=q_1+iq_2$.
With the ansatz $\eta_{1,2}=\beta_{1,2}(z)\eta(x)$,
the equations reduce to %
\begin{itemize}
\item[(1)]
the four-dimensional Killing spinor equation for $\eta(x)$,
\be
\label{k4}
2\wh D_m\eta+i\la g \wh e_m{}^a\si_a\etabar \ =\ 0, 
\ee
where $g\in \mb C$,
$ g^\ast = q_{12}\beta_1^2-q_{12}^\ast\beta_2^2+2q_3\beta_1\beta_2$
and $gg^\ast=K^2$;
\item[(2)]
equations for the fermionic warp factors,
\be
\label{b12}
2\mc{B}^\prime+\la(\vec p\cdot\vec \si)\mc{B}=0, \quad
\mc{B}\equiv\left(\bma \beta_1 \\ \beta_2\ema\right) \quad
\Leftrightarrow \quad
\left\{\bma
2\beta_1^\prime+\la(q_3\beta_1-q_{12}^\ast\beta_2)=0, \\[2mm]
2\beta_2^\prime-\la(q_3\beta_2+q_{12}\beta_1)=0,\ema\right.
\ee
where $\vec p=(-q_1, -q_2, q_3)$; these equations
ensure that $g$ is a constant;
\item[(3)] 
a relation between the bosonic and fermionic warp factors,
\be
\label{abb}
a(z)=\mc{B}^\dagger\mc{B}=\beta_1\beta_1^\ast+\beta_2\beta_2^\ast.
\ee
\end{itemize}

The Killing spinor equation (\ref{k4}) has solutions in the
\AdS\ and \Mink\ backgrounds.
Equations (\ref{b12}) for the fermionic warp factors are easy to solve,
\be
\beta_i(z)=U(z)_i{}^j\beta_j(0), \quad U(z)=\exp
\left(-\frac{\la}{2}(\vec p\cdot\vec \si)z\right).
\ee
The solution shows that the Killing spinor is {\it twisted}.
(Note that $U(z)$ commutes with the gauged 
$U_{\vec q}(1)\subset SU(2)_R$, but $U(z)\notin SU(2)_R$.)

Therefore we see that the Killing spinor equations can be solved 
in {\it all} the (static) warped backgrounds. However, a true Killing
spinor must also satisfy the boundary conditions (\ref{etabc}),
\be
\beta_2(0)=\al_0\beta_1(0), \quad
\beta_2(\pi R)=\al_\pi\beta_1(\pi R).
\label{ksbc}
\ee
Let us fix $q_3=1$ to simplify the discussion. (Other choices 
of $\vec q$ can be obtained by an $SU(2)_R$ rotation.)  Then 
\be
\beta_1(z)=\exp(-\frac{\la}{2}z)\beta_1(0), \quad
\beta_2(z)=\exp(+\frac{\la}{2}z)\beta_2(0)
\ee
and the boundary conditions (\ref{ksbc}) require
\be
\label{alrel}
\al_\pi=\al_0 \exp(\la\pi R).
\ee
When this condition is not satisfied, there is no Killing spinor and
supersymmetry is spontaneously broken.

Consider now the set of boundary conditions consistent with local
supersymmetry.  For $q_3=1$, Eqs.~(\ref{fff}) and (\ref{fbc2}) imply
\be
\label{alq3}
|\al_0|^2=\frac{\la-\la_0}{\la+\la_0}, \qquad
|\al_\pi|^2=\frac{\la-\la_\pi}{\la+\la_\pi}.
\ee
(The same relations follow from Eq.~(\ref{abb}).)
Equation (\ref{rrr}) then implies
\be
|\al_\pi|=|\al_0|\exp(\la\pi R).
\ee
This relates the magnitudes $|\al_0|$ and $|\al_\pi|$, but
does not determine the phases $\al_0=|\al_0|e^{i\vp_0}$ and 
$\al_\pi=|\al_\pi|e^{i\vp_\pi}$. If the phases are the same,
$\vp_0=\vp_\pi$ ($+2\pi n$, $n\in\mb{Z}$), Eq.~(\ref{alrel}) is
satisfied and supersymmetry is not broken.  
Otherwise, Eq.~(\ref{alrel}) cannot be satisfied;
there is no Killing spinor and supersymmetry is spontaneously broken
by the twisted boundary conditions.

\subsection{Shift in the Kaluza-Klein spectrum}
\label{sec kk}

We conclude this section with a simple example which illustrates how
the phases $\vp_0$ and $\vp_\pi$ shift the masses for the fermionic
Kaluza-Klein modes.

We set $q_3=1$, and take the ansatz $\psi_{n1,2}=b_{1,2}(z)\psi_n(x)$,
where $\psi_n(x)$ satisfies the four-dimensional gravitino equation with
mass parameter $m\in\mb{R}$,
\be
\wh\eps^{\,mpnk}\wh\si_p\wh D_n\ov\psi_k+2m\wh\si^{mn}\psi_n=0.
\ee
The five-dimensional fermionic equations of motion give rise to the
following equations for the warp factors $b_{1,2}(z)\in\mb{C}$:
\be
\label{kkeq}
b_1^\prime+(\frac{3\la}{2}+\frac{a^\prime}{a})b_1=\frac{m}{a}b_2^\ast,
\quad
b_2^\prime+(-\frac{3\la}{2}+\frac{a^\prime}{a})b_2=-\frac{m}{a}b_1^\ast.
\ee
The mass quantization follows from the boundary conditions,
\be
b_2(0)=\al_0 b_1(0), \quad b_2(\pi R)=\al_\pi b_1(\pi R).
\ee

For the \AdS\ case, the bosonic warp factor is $a(z)=K\cosh(\la z-c_0)$. 
The brane tensions are given by
\be
\la_0=\la\tanh(c_0), \quad \la_\pi=-\la\tanh(\la\pi R-c_0).
\ee
As discussed above,
local supersymmetry fixes the absolute values of $\al_0$ and $\al_\pi$, 
but allows arbitrary complex phases $\varphi_0$ and $\varphi_\pi$:
\be
\al_0=e^{-c_0}e^{i\phi_0}, \quad \al_\pi=e^{\la\pi R-c_0}e^{i\phi_\pi}.
\ee
For this example, we follow Ref.~\cite{lm} and take $\la_0=-\la_\pi$
($T_0=T_\pi$).  In this case, $c_0=\half\la\pi R$.

We now assume that $\la\pi R\ll 1$.  Equations (\ref{kkeq})
simplify as follows,
\be
\label{simp}
b_1^\prime(y)+\frac{3}{2}b_1=M b_2^\ast, \quad
b_2^\prime(y)-\frac{3}{2}b_2=-M b_1^\ast, 
\ee
where $y=\la z-c_0$ and $M=m/(\la K)$.
($M=1$ corresponds to the massless mode in the \AdS\ space
with cosmological constant $\Lambda_4=-3\la^2 K^2$.)
Equations~(\ref{simp}) are easy to solve; the solutions fall into two
classes, depending on whether $M<1.5$ or $M>1.5$.  For
$M\gg 1$, the solutions are
\be
b_1(y)=A\cos(M y)+B\sin(M y), \quad
b_2(y)=B^\ast\cos(M y)-A^\ast\sin(M y).
\ee
With our assumptions, we find $K=1$, $\al_0=e^{i\varphi_0}$ and
$\al_\pi=e^{i\varphi_\pi}$.  Setting $A=A_0 e^{i\vartheta_1}$
and $B=B_0 e^{i\vartheta_2}$, we cast the boundary conditions
into the following form,
\bea
&& \sin(\frac{\vp_0+\vp_\pi}{2}+\vt_1+\vt_2)\cos(\frac{\vp_0-\vp_\pi}{2})=0,
\nn\\
&& \sin(\frac{\vp_0+\vp_\pi}{2}+\vt_1+\vt_2)\sin(\frac{\vp_0-\vp_\pi}{2})=
\sin(m\pi R)\cos(\vt_2-\vt_1),
\nn\\
&& \cos(\frac{\vp_0+\vp_\pi}{2}+\vt_1+\vt_2)\sin(\frac{\vp_0-\vp_\pi}{2})=
\sin(m\pi R)\sin(\vt_2-\vt_1),
\nn\\
&&\frac{A_0}{B_0}=
\frac{\cos(\vp_0+\vt_1+\vt_2)+\sin(m\pi R)\cos(\vt_2-\vt_1)}
     {\cos(m\pi R)}.
\eea
The solution, valid for any $\vp_0$ and $\vp_\pi$, is
\be
\frac{A_0}{B_0}=1, \quad
\vt_2-\vt_1=\frac{\pi}{2}, \quad
\vt_1+\vt_2=-\frac{\vp_0+\vp_\pi}{2}+\pi j, \quad
m\pi R=\pi j+\frac{\vp_0-\vp_\pi}{2}.
\ee
The Kaluza-Klein mass for the $j$'th gravitino mode is given by
\be
m_j=\frac{j}{R}+\frac{\vp_0-\vp_\pi}{2\pi R}, \quad j\in\mb{Z}.
\ee
This formula is valid for $m_j\gg\la K\approx\la$. 
Because $\la R\ll 1$, the condition is satisfied for all modes, except,
perhaps, the lightest one.  Note that the mass 
shift depends only on the {\it phase difference}.  This must 
always be true because one phase can be absorbed by a field
redefinition.

\section{Lifting ``upstairs''}

\subsection{Introduction}
In this section we show how to lift the previous construction to the 
``upstairs'' picture, in which the fifth dimension is the circle
$S^1$ or the line $\mathbb{R}$.  Lifting to a covering space
brings some technical and conceptual advantages.  For example,
a manifold without boundary allows one to neglect total derivatives,
while a simply connected manifold allows one to avoid
multi-valued fields.

\subsection{General procedure}

We start by describing our general lifting procedure, 
using a symmetry that is broken when it acts
only on fields, but is intact when the parameters also
rotate.  (For the case at hand, we use the $SU(2)_R$
symmetry, broken by the gauged $U(1)$ subgroup.)  The
basic idea is to combine a broken symmetry transformation
with a group motion on the covering space, choosing
appropriate parameters on the different domains.

To see how this works, let us consider an action for fields
$\Phi(x)$ on a space $\mc{M}$ with a set of parameters $Q$:
\be
S\ =\ \int_{\mc{M}}dx\mc{L}[\Phi(x), Q].
\ee
We assume that $S$ is invariant under $G=\{\wh g_i\}$, a
discrete group of transformations that acts on $\Phi$, $\mc{M}$
and $Q$,
\be
\wh g_i S = \int_{\wh g_i(\mc{M})} dx^\prime 
\mc{L}[\Phi^\prime(x^\prime), \wh g_i Q]
= \int_{\mc{M}}dx\mc{L}[\Phi(x), Q] = S.
\ee
(Here $x^\prime=\wh g_i x$, $\Phi^\prime=\wh g_i \Phi$;
$\wh g_i$ is a conventional symmetry if $\wh g_i Q = Q$.)  
We also assume that the group action splits $\mc{M}$ into a
set of disjoint subspaces, such that
\be
\mc{M}\ =\ \cup\mc{M}_i, \quad \mc{M}_i\ =\ \wh g_i \mc{M}_0,
\ee
where $\mc{M}_0\simeq\mc{M}/G$ is the fundamental domain.

We now wish to construct an action that is invariant under $G$,
with the group acting on $\Phi$ and $\mc{M}$, but {\it not} on
the parameters $Q$.  Since
\be
\wh g_i\int_{\mc{M}_0} dx\mc{L}[\Phi(x), Q] =
\int_{\wh g_i(\mc{M}_0)} dx^\prime\mc{L}[\Phi^\prime(x^\prime), Q] =
\int_{\mc{M}_i} dx\mc{L}[\Phi(x),  {\wh g_i}^{-1} Q],
\ee
the action
\be
\label{lifting}
\wt S\ =\ \sum_i\int_{\mc{M}_i}\mc{L}[\Phi(x),Q_i],
\ee
with $Q_i= {\wh g_i}^{-1} Q$,
describes a $G$-invariant theory on the covering space $\mc{M}$. 

Now suppose that we restrict the fields $\Phi(x)$ to $G$-invariant
configurations
\be
\wh g_i \Phi(\wh g_i x)\ =\ \Phi(x).
\ee
In this case the theory on the covering space is equivalent to
the theory on the fundamental domain, $\mc{M}_0$.  Only the
fundamental domain $\mc{M}_0$ is physical; all other domains
are its ``mirror images.''  The space $\mc{M}_0 \simeq \mc{M}/G$ 
is, in general, an orbifold.

\subsubsection{Example:  Supergravity on $S^1/\mathbb{Z}_2$}
\label{sec z2}

To illustrate the lifting procedure, we lift supergravity from
the fundamental domain $\mc{M}_0 = [0, \pi R]$ to its
covering space $\mc{M} = S^1$.  Even
in this simplest case, the lifting relies on a broken symmetry.  
This is the origin of the so-called ``odd bulk mass term'' in the
supersymmetric Randall-Sundrum scenario.

We use the discrete group $G=\mathbb{Z}_2$,
generated by the parity transformation $\cP$.  The group
acts on the fifth coordinate
($x^5 = z$) and on the fields and supersymmetry parameters
according to $\cP z = -z$ and $\cP \Phi=P(\Phi)\Phi$, where
\bea
\label{parity}
P(e_m^a, e_5^\hf, B_5, \psi_{m1}, \psi_{52}, \eta_1) &=& +1, \nn\\
P(e_m^\hf, e_5^a, B_m, \psi_{m2}, \psi_{51}, \eta_2) &=& -1.
\eea

When $q_3\neq 0$ and $\mc{M} = S^1$ ($z\in [-\pi R, \pi R]$), the
action (\ref{bulk}) is not invariant under
$\cP$.  However, the action {\it is} invariant if $\cP$ acts on $\vec q$
as follows,
\be
\cP (q_1, q_2, q_3) \ =\  (q_1, q_2, -q_3).
\ee
This allows us to construct an invariant action
following the procedure described 
in the previous section.  The lifted action
is just the bulk action (\ref{bulk}), with $q_3$ having a
different sign on each side of the circle:
\be
\label{z2bulk}
\wt S_{bulk} = \int_{-\pi R}^0\!dz \int d^4\!x \mc{L}_{bulk}[-q_3] +
               \int_0^{\pi R}\!dz \int d^4\!x \mc{L}_{bulk}[q_3].
\ee
The parameter $\vec q$ multiplies a gravitino bilinear,
so $q_3$ is responsible for the ``odd bulk mass term.''

In Ref.~\cite{bb1} we found that warped backgrounds
also require a brane action,
\bea
\label{z2brane}
S_{brane}
&=&\int d^5\!x e_4
(-6\la_0-2\al_0(\psi_{m1}\si^{mn}\psi_{n1}+h.c.))\ \da(z)
\nn\\
&+&\int d^5\!x e_4
(+6\la_\pi+2\al_\pi(\psi_{m1}\si^{mn}\psi_{n1}+h.c.))\ \da(z-\pi R),
\eea
for arbitrary $\vec q$ in the bulk.  The brane action
gives rise to jumps and cusps in the fields.  The total bulk-plus-brane
action, the sum of (\ref{z2bulk}) and (\ref{z2brane}), is
supersymmetric, provided $\la_{0,\pi}=f(\al_{0,\pi},\vec q\,)\la$,
where $f(\al,\vec q\,)$ is defined in Eq.~(\ref{fff}).  In
general, the supersymmetry transformation for $\psi_{52}$
must also be modified,
\be
\da\psi_{52} =\da\psi_{52}\Big\vert_{\rm old}
-4(\alpha_0\da(z)+\alpha_\pi\da(z-\pi R))\eta_1 ,
\ee
in which case the supersymmetry algebra closes and $\da\psi_{52}$ 
is finite on the branes (see Ref.~\cite{bb1}).

The $S^1/\mathbb{Z}_2$ approach enjoys a few technical and
conceptual advantages:
\begin{enumerate}
\item[(1)]
Total derivatives can be dropped when checking invariance
of the action under supersymmetry.
\item[(2)]
The boundary conditions given in Section \ref{babc} follow from
the action principle.  (The jumps are induced by the brane action;
together with
the parity assignments they imply the boundary conditions.)
\item[(3)]
The parameters $\la_{0,\pi}$ and $\al_{0,\pi}$ acquire clear physical 
interpretations as brane tensions and brane localized mass terms,
respectively.
\end{enumerate}
We emphasize, though, that all physical statements established
on the fundamental domain remain unchanged on the covering space.
The construction introduces mirror images of spacetime that do
not change the boundary conditions.


\subsection{Scherk-Schwarz twisting}

In this section, we use a generalization of the previous construction
to lift  $\mc{M}_0=[0,\pi R]$ to $\mc{M}=\mathbb{R}$, with periodic
bosonic fields and twisted fermionic fields, along the lines of
Scherk and Schwarz \cite{ss}.  We assume that the bosonic background
is warped.  The warp factor $a(z)$ does not satisfy $a(0)=a(\pi R)$,
so we cannot use a simple translation of the bosonic fields.
Instead, we use a set of parity reflections to make $a(z)$ continuous
along $\mathbb{R}$.  We construct twisted embeddings by choosing
different parity operators at the ends of the interval $[0,\pi R]$.

\subsubsection{Twisted parity}

In Section \ref{sec z2} we defined the standard parity transformation
$\cP$.  We now consider a more general parity transformation
$\bP$, one that includes an $SU(2)_R$ rotation of fermions.
Its action on the fifth coordinate, bosonic fields and fermionic fields is,
respectively,
\be
\label{mcptr}
\bP z = -z, \quad
\bP \phi = P(\phi) \phi, \quad
\bP \Psi = U Z \Psi,
\ee
where $U\in SU(2)_R$, $Z=\si_3=\left(\bma 1 & 0\\0 & -1 \ema\right)$
and $\Psi=\left(\bma \psi_{m1} \\ \psi_{m2} \ema\right)$ (and likewise
for $\psi_{5}$ and $\eta$).  The requirement $\bP\bP=
{\bf 1}$ implies
\be
\label{uz}
(U Z)^2\ =\ 1 \quad \Leftrightarrow \quad Z U Z \ =\ U^{-1}.
\ee
We denote by $\mc S$ the set of $U$ satisfying this condition.
Note that $\mc S$ is {\it not} a group.

A general element $U\in SU(2)_R$ can be written as follows,
\be
U=u_0+ i\vec u\cdot\vec\si =
\left(\bma u_{03} & u_{21} \\ 
-u_{21}^\ast & u_{03}^\ast \ema\right), \quad
|u_{03}|^2+|u_{21}|^2=1.
\ee
(We use notation $u_{ab}=u_a+i u_b$.)  For the set $\mc S$, we find
\be
U\in \mc S \quad \Leftrightarrow \quad 
\si_3 U \si_3 = U^{-1} = U^\dagger
\quad \Leftrightarrow \quad u_3=0.
\ee
Taking $u_0=\cos\theta$ and $u_{21}=e^{i\phi}\sin\theta$, we
can parametrize $\mc S$ as follows,
\be
U(\theta, \phi)= 
\left(\bma \cos\theta & e^{i\phi}\sin\theta \\
           -e^{-i\phi}\sin\theta & \cos\theta \ema\right) .
\ee

The $SU(2)_R$ symmetry is broken by the vector $\vec q$. 
Only transformations in $U_{\vec q}(1)$,
\be
U=\exp(i\omega\vec p\cdot\vec\si) =
\cos\omega + i(\vec p\cdot\vec\si)\sin\omega,
\ee
with $\vec p=(-q_1,-q_2,q_3)$, leave the action invariant.\footnote{
When $q_3\neq 0$, $U=\pm 1$ are the only {\it unbroken} symmetry 
transformations that satisfy Eq.~(\ref{uz}).  These are the transformations
that underlie the ``flipped twisting'' of Refs.~\cite{gp,bfl,lm}.}
Invariance under {\it any} $U\in SU(2)_R$ is achieved if we
rotate $\vec q$,
\be
{\bf U}\vec q \ =\  \vec q^{\;\prime}, \quad 
(\vec p^{\;\prime}\cdot\vec\si) \ =\ U(\vec p\cdot\vec\si)U^\dagger .
\ee
Since $\bP={\bf U} \cP$, this determines the action of
$\bP$ on $\vec q$. 

\subsubsection{Bulk action}

The bulk action depends on fields lifted from the fundamental
domain $\mc{M}_0=[0,\pi R]$ to the covering space $\mc{M}=
\mathbb{R}$.  We use {\it different} parity identifications 
across the branes at $z=0$ and $z=\pi R$,
\bea
\Psi(-z)&=&U_0 Z\Psi(z), \nonumber\\
\Psi(\pi R-z)&=&U_1 Z\Psi(\pi R+z),
\eea
with $U_{0,1}\in \mc S$.  Together, these imply
\be
\Psi(z+2\pi R)\ =\ U_{\Delta}\Psi(z), \quad 
U_{\Delta}=U_1 U_0^{-1}.
\ee
In this construction, the bosonic fields are periodic under shifts
by $2\pi R$, but the fermionic fields twist by $U_{\Delta} \in SU(2)_R$.

The twists $U_0$ and $U_1$ define a lifting of the
bulk action (\ref{bulk}) from $[0,\pi R]$ to $\mathbb{R}$
as in (\ref{lifting}).  The parameters $\vec q_{(n)}=
(q_1,q_2,q_3)$ on the intervals $\mc{M}_n=\big(n\pi R, (n+1)\pi R\big)$
are given by
\be
\label{qn}
\left(\bma -q_3 & q_{12}^\ast \\ 
           q_{12} & q_3 \ema\right){\Bigg\vert_{\mc{M}_n}} =
\left\{\bma\dst
(U_\Delta)^{-k} 
\left(\bma -q_3 & q_{12}^\ast \\ q_{12} & q_3 \ema\right)
(U_\Delta)^k,  &n=2k, \\
(U_\Delta)^{-k} U_0^{-1}
\left(\bma q_3 & q_{12}^\ast \\ q_{12} & -q_3 \ema\right)
U_0 (U_\Delta)^k,  &n=2k-1.
\ema\right.
\ee
With this choice of parameters, the bulk action is invariant under
translations by $2\pi R$, provided the fermions transform as
\be
\Psi(z) \rightarrow \Psi^\prime(z+2\pi R)=U_\Delta \Psi(z),
\ee
and the bosonic fields are trivially shifted, $\phi^\prime(z+2\pi R)=\phi(z)$.
Note that the parameters $\vec q_{(n)}$ are {\it not} rotated.

\subsubsection{Brane action}

The parities for the mirror images of the two physical branes are given 
by\footnote{
With $U_{0,1}\in \mc S$, one can show that $U_n\in \mc S$,
even though $U_\Delta \notin \mc S$ in general. 
Any power of $U_n$ is well-defined and belongs to $\mc S$
because $U(\theta,\phi)^w=U(w\theta,\phi)$.
}
\be
\label{nparity}
\Psi(n\pi R-z)\ =\ U_n Z\Psi(n\pi R+z), \quad
U_n=(U_{\Delta})^n U_0 .
\ee
Note that the two-component spinors $\psi_{m1}$ and $\psi_{m2}$
are {\it not} of definite parity.  However, after the redefinitions
%
\be
\label{hatted}
\wh\Psi^{(n)}(z) \ =\  U_n^{-\frac{1}{2}} \Psi(z), 
\ee
the fields $\wh\psi_{m1}^{(n)}$ and $\wh\psi_{m2}^{(n)}$ are
even and odd, respectively, across the $n$'th brane, because
\be
\label{whnpar}
\wh\Psi^{(n)}(n\pi R-z)\ =\ Z\wh\Psi^{(n)}(n\pi R+z).
\ee
Using the parametrization $U_n=U(\theta_n,\phi_n)$, we find
\bea
\label{hat12}
&&\wh\psi_{m1}^{(n)}\ =\ \psi_{m1}\cos(\half\theta_n)
-\psi_{m2}e^{i\phi_n}\sin(\half\theta_n),\nonumber \\
&&\wh\psi_{m2}^{(n)}\ =\ \psi_{m1}e^{-i\phi_n}\sin(\half\theta_n)
+\psi_{m2}\cos(\half\theta_n).
\eea

As in Ref.~\cite{bb1}, the odd fermionic fields can be discontinuous
across the branes.  The jumps are determined by brane actions for
the fermi fields.  Using the fields of definite parity, we can simply
copy the results of Ref.~\cite{bb1} (see Section \ref{sec z2}).  The
complete brane action is just the sum of the individual brane actions,
\be
\label{nbrac}
S_\text{brane} = \sum\limits_{n\in\mathbb{Z}}\int\!d^5x e_4 \left(
-6\la_n -2 \wh\al_n (\wh\psi_{m1}^{(n)}\si^{mn}\wh\psi_{n1}^{(n)} + h.c.)
\right)\da(z-n\pi R).
\ee
It is easy to rewrite this action in terms of the original fields $\psi_{m1}$
and $\psi_{m2}$, using Eq.~(\ref{hat12}).

The brane action implies the boundary conditions 
\be
\wh\psi_{m2}^{(n)}=\pm\wh\al_n\wh\psi_{m1}^{(n)}, \quad
\om_{ma\hf}=\pm\la_n e_{ma},
\ee
with all fields evaluated at $z_n^{\pm}=n\pi R\pm 0$.  In terms of
the original fields $\psi_{m1}$ and $\psi_{m2}$, 
the boundary conditions are
\be
\psi_{m2}(z_n^{\pm})=\al_n^{\pm}\psi_{m1}(z_n^{\pm}), \quad
\al_n^{\pm} =\frac{\pm\wh\al_n-e^{-i\phi_n}\tan(\half\theta_n)}
{1\pm\wh\al_n e^{i\phi_n}\tan(\half\theta_n)}.
\ee
Local supersymmetry requires $\la_n=f(\al_n^+,{\vec q}_{(n)})\la$,
where $f(\al,\vec q\,)$ is defined in Eq.~(\ref{fff}) and ${\vec q}_{(n)}$
is given in Eq.~(\ref{qn}).

\subsubsection{Jumps, twists and boundary conditions}

In terms of the spinors 
$\Psi=\left(\bma \psi_{m1}, \psi_{m2}\ema\right)^T$, 
the discontinuities across the branes can be parametrized
as
\be
\wh \Psi^{(n)}(n\pi R-0)\ =\ \ov U_n \wh\Psi^{(n)}(n\pi R+0),
\ee
where $\ov U_n\in \mc S$.  This gives rise to the following
boundary conditions for the fermi fields,
\be
\Big(\ov U_n^{\pm1} - Z\Big)\wh\Psi^{(n)}(n\pi R \pm 0)=0.
\label{hatjump}
\ee
In general, the condition $(U-Z)\Psi=0$ implies 
$\psi_{m2}=\mc{A}(U)\psi_{m1}$, where
\be
\mc{A}(U)\equiv 
\frac{1-u_{03}}{u_{21}} = \frac{u_{21}^\ast}{1+u_{03}^\ast}.
\ee
(Consistency requires $u_3=0$, so $U\in\mc{S}$.)
For the hatted fields, Eq. (\ref{hatjump}) implies
\be
\wh\al_n=\mc{A}(\ov U_n).
\ee
Analogously, in terms of the original fields, we find
\be
\left(U_n^{-\frac{1}{2}} \ov U_n^{\pm1} U_n^{-\frac{1}{2}} - Z\right)
\Psi(n\pi R \pm 0)=0,
\ee
and therefore
\be
\al_n^\pm=\mc{A}(U_n^{-\frac{1}{2}} \ov U_n^{\pm1} U_n^{-\frac{1}{2}}).
\ee
Note that $\mc{A}(U)$ determines $U$ uniquely. 
For $U=U(\theta, \phi)$, the parameters
$\theta$ and $\phi$ can be read from $\mc{A}(U)$ 
as follows:
\be
\mc{A}(U)=e^{-i\phi}\tan(\half\theta).
\ee

Because there are just two physical branes, only quantities with 
$n=0$ and $1$ are independent. The jump matrices $\ov U_n$ obey the relations
\be
U^{\frac{1}{2}}_{n}\ov U_n U^{-\frac{1}{2}}_{n} =\left\{\dst\bma
U_\Delta^k\ov U_0 U_\Delta^{-k},& n=2k, \\[2mm]
U_\Delta^k\ov U_1 U_\Delta^{-k},& n=2k+1.
\ema\right.
\ee
For a fixed set of twists ($U_0$, $U_1$), the matrices $\ov
U_0$ and $\ov U_1$ are determined by the boundary
conditions on the fundamental interval,
\be
\psi_{m2}(+0)=\al_0\psi_{m1}(+0), \quad
\psi_{m2}(\pi R-0)=\al_\pi\psi_{m1}(\pi R-0).
\ee
We find 
\be
\al_0=\al_0^+=\mc{A}(U_0^{-\frac{1}{2}}\ov U_0 U_0^{-\frac{1}{2}}), \quad
\al_\pi=\al_1^-=\mc{A}(U_1^{-\frac{1}{2}}\ov U_1^{-1} U_1^{-\frac{1}{2}}).
\ee
There is freedom to choose the twists and jumps, so long as
the boundary conditions on the fundamental
interval remain unchanged (see also
Refs.~\cite{bff, lm2}).
In particular, we can trade twists for jumps, and vice versa:
\begin{enumerate}
\item
Continuous twisted fields (no jumps, $\wh\al_n=0$): 
$\ov U_0=1$, $\ov U_1=1$
\be
\al_0=\mc{A}(U_0^{-1})=-e^{-i\phi_0}\tan(\half\theta_0), \quad
\al_\pi=\mc{A}(U_1^{-1})=-e^{-i\phi_1}\tan(\half\theta_1).
\ee
\item
Jumping periodic fields (no twist, $U_\Delta=1$): $U_0=1$, $U_1=1$
\be
\al_0=\mc{A}(\ov U_0)=e^{-i\ov\phi_0}\tan(\half\ov\theta_0), \quad
\al_\pi=\mc{A}(\ov U_1^{-1})=-e^{-i\ov\phi_1}\tan(\half\ov\theta_1).
\ee
\end{enumerate}
The physics is the same for each case.

\subsubsection{Example}

In Section \ref{sec kk} we derived the Kaluza-Klein spectrum for the
gravitini $\psi_{m1,2}=b_{1,2}^{(j)}(z)\psi^{(j)}_m(x)$, in the case
where
\be
q_3=1, \quad \la_1=-\la_2, \quad \la\pi R \ll 1, \quad m_j \gg \la.
\ee
We set $\al_0=e^{i\vp_0}$, $\al_\pi=e^{i\vp_\pi}$, and found that
the warp factors on the fundamental interval, $z\in[0,\pi R]$, are
given by
\bea
\label{fsol}
&&b_1^{(j)}(z) = A_0^{(j)}
\exp\{+i(m_{j} z-\frac{\vp_0}{2})\},
\nn\\
&&b_2^{(j)}(z) = A_0^{(j)}
\exp\{-i(m_{j} z-\frac{\vp_0}{2})\},
\eea%
where 
\be
m_{j}=\frac{j}{R}+\frac{\vp_0-\vp_\pi}{2\pi R}
\ee
is the Kaluza-Klein mass for the $j$'th mode and
$A_0^{(j)}$ is a (complex) normalization constant.

Let us now apply the lifting procedure described above.
For the twisted lifting without jumps, with $\ov U_0=\ov U_1=1$, we
deduce $U_0=U(\theta_0,\phi_0)=U(-\frac{\pi}{2},-\vp_0)$ and
$U_1=U(\theta_1,\phi_1)=U(-\frac{\pi}{2},-\vp_\pi)$, which
implies
\be
U_0= \left(\bma 0 & -e^{-i\vp_0} \\ e^{i\vp_0} & 0 
\ema\right), \quad
U_\Delta= \left(\bma e^{i(\vp_0-\vp_\pi)} & 0 \\ 0 & e^{i(\vp_\pi-\vp_0)} 
\ema\right).
\ee
From Eq.~(\ref{qn}) we find $\vec q_{(n)}=\vec q =(0, 0, 1)$.
Lifting the fields from $[0,\pi R]$ to $\mb{R}$, using
\be
\Psi(-z)=U_0 Z \Psi(z), \quad
\Psi(z+2n\pi R)=(U_\Delta)^n \Psi(z),
\ee
it is not hard to show that the $b_{1,2}^{(j)}(z)$ are given by Eq.~(\ref{fsol})
{\it for all} $z\in \mb{R}$.  In particular, they are continuous, so
$\ov U_n=1$ and $\wh\al_n=0$. However, because of the
nontrivial twists $U_n$,
\be
U_n=U(\theta_n, \phi_n)=U(-\tst\frac{\pi}{2}, n(\vp_0-\vp_\pi)-\vp_0),
\ee
the $\al_n^\pm$ are nonzero, and given by 
$\al_n^\pm=\mc{A}(U_n^{-1})=e^{-i\phi_n}$.

For the periodic lifting with jumping fields, in which case 
$U_0=U_1=1$, we apply
\be
\Psi(-z)= Z \Psi(z), \quad
\Psi(z+2n\pi R)=\Psi(z).
\ee
Using Eq.~(\ref{fsol}) for $z\in[0,\pi R]$, we can write explicit
expressions for the lifted fields for any $z\in\mb{R}$.  For the jump
matrices, we find
\be
\ov U_0=\ov U_{2k}=U(\tst\frac{\pi}{2}, -\vp_0), \quad
\ov U_\pi=\ov U_{2k-1}=U(-\tst\frac{\pi}{2}, -\vp_\pi),
\ee
so $\wh\al_{2k}=e^{i\vp_0}$ and $\wh\al_{2k-1}=-e^{i\vp_\pi}$.
Since there are no twists, we have $\al_n^\pm=\pm\wh\al_n$,
as well as $\vec q_{2k}=(0,0,1)$ and $\vec q_{2k-1}=(0,0,-1)$.
For this lifting, the warp  factors $b_{1,2}^{(j)}(z)$ are of definite parity:
$b_1^{(j)}(z)$ is even and continuous, while $b_2^{(j)}(z)$ is odd
and jumping.  The two cases for this example are illustrated in
Figure 1.

\FIGURE[t]{
\epsfig{file=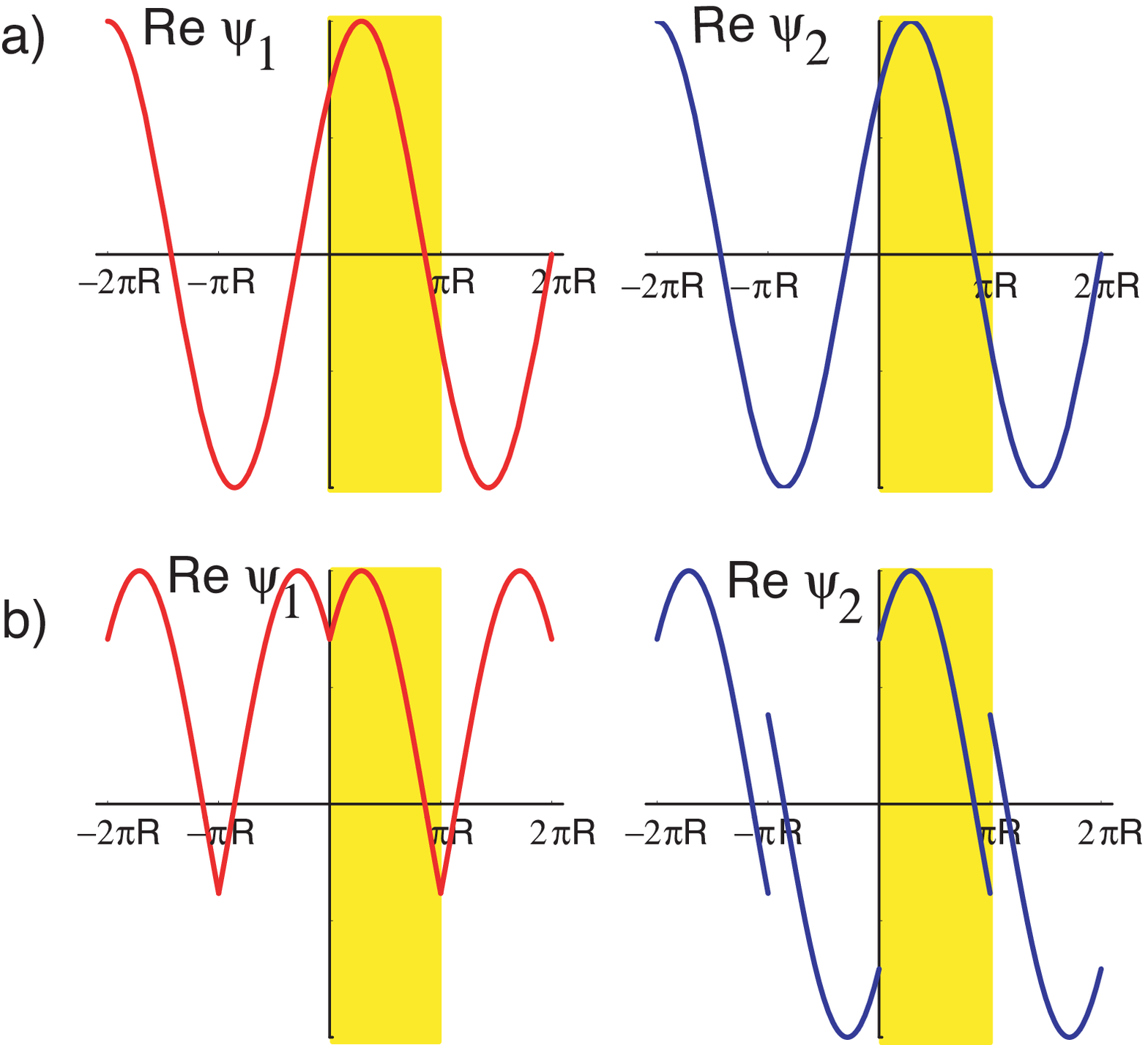,width=4.3in} 
\caption{Two gravitino modes, from the example in section 3.3.5.
The physics is determined by the fundamental domain, shaded. 
a)  A continuous, but not $2\pi R$-periodic, lifting to $\mb{R}$.  
b)  A $2\pi R$-periodic, but not continuous, lifting to $\mb{R}$.  
With this lifting, $\psi_1$ is even, and $\psi_2$ is odd.}
}

\subsection{Summary}

In this paper, we studied supergravity on a slice of
{\it AdS$_5$}.  Using the ``downstairs'' picture, we presented 
locally supersymmetric boundary conditions that support
warped backgrounds.  For every bosonic background,
we found a set of fermionic boundary conditions
($\al_0, \al_\pi$) that preserve global supersymmetry.
For every \AdS\ background, we also found a
continuous family of boundary conditions that
spontaneously break global supersymmetry.
The set is parametrized by the phase difference 
$\Delta\vp=\vp_0-\vp_\pi$, where the $\vp_i$ are the
phases of $\al_i$ when $q_3=1$.

The ``flipped'' boundary conditions of Refs.~\cite{lm,bfl}
correspond to the case $\Delta\vp=\pi$.  In this
paper we generalized the flip to a continuous set
of supersymmetry-breaking boundary conditions.
The mass shift for the Kaluza-Klein fermionic modes
depends on $\Delta\vp$ and can be turned continuously
on and off. 

The spontaneous supersymmetry breaking discussed here
occurs only in \AdS\ backgrounds.  When the brane
tensions are tuned, as in the original Randall-Sundrum
scenario, the boundary conditions are fixed  {\it uniquely}.  
(For example, Eq.~(\ref{alq3}) requires that $\al_0=0$
when $q_3=1$ and $\la_0=\la$.)  Furthermore,
the boundary conditions are such that a Killing spinor
always exists, and supersymmetry is not broken.  This
agrees with the conclusion of Ref.~\cite{hall}.

The ``upstairs'' picture is based on a lifting of the physical
(fundamental) domain.  The lifting uses a {\it broken}
symmetry group $G$, a combination of rotations on the fields and
motions on the covering space.  The symmetry is such that it can
be restored by rotating the parameters as spurions. It can also be
restored by choosing different parameters on the different domains,
consistent with the group $G$, which maps one domain to
another.  We take the second approach, and choose the parameters
so that the lifted theory is $G$-invariant ({\it without} rotating the
parameters).

The lifting from $[0,\pi R]$ to $S^1$ uses $G=\mb{Z}_2$,
generated by the reflection $x^5\rightarrow -x^5$ together
with a parity transformation on the fields.  The Scherk-Schwarz
lifting to $\mb{R}$ uses a twisted parity on the fermionic fields.
In this case $G=\mb{Z}_2\times\mb{Z}$ (a semi-direct product).

In each case, the physics is uniquely specified by the boundary
conditions on the fundamental domain. The lifting, however, is
not unique.  There is freedom to choose the twist and jump
parameters.  In fact, one can construct twisted lifting even
when the boundary conditions ($\al_0,\al_\pi$) do not break
supersymmetry.

\vspace{0.2in}

We would like to thank A.\ Delgado, F.\ Feruglio, M.\ Quiros,
M.\ Redi and F.\ Zwirner for discussions.  This work was
supported in part by the U.S. National Science Foundation, grant
NSF-PHY-9970781.


\end{document}